\documentclass[useAMS,usenatbib,letterpaper]{mn2e}
\usepackage{amsmath}                       

\usepackage{amsfonts}
\usepackage{amssymb}
\usepackage{graphicx}
\usepackage{float}
\usepackage{morefloats}
\usepackage{listings}
\usepackage{color,hyperref}
\definecolor{linkcolor}{rgb}{0,0,0.25}
\hypersetup{
  colorlinks=true,        
  linkcolor=linkcolor,    
  citecolor=linkcolor,    
  filecolor=linkcolor,    
  urlcolor=linkcolor      
}

\newcommand{\kms}{\ensuremath{\mathrm{km}~\mathrm{s}^{-1}}}

\newcommand{\pc}{\ensuremath{\mathrm{pc}}}
\newcommand{\kpc}{\ensuremath{\mathrm{kpc}}}
\newcommand{\rvr}{\ensuremath{\Delta R-v_{\mathrm{R}}}}
\newcommand{\zvz}{\ensuremath{z-v_z}}

\definecolor{kvjcolor}{rgb}{0.75,0,0}

\definecolor{apwcolor}{rgb}{0,0.75,0.75}

\definecolor{jhcolor}{rgb}{0,0,0.75}

\definecolor{edcolor}{rgb}{0,0.75,0.0}

\title[Radial phase spirals]
{Radial phase spirals in the Solar neighbourhood}
\author[J. A. S. Hunt et al.]
  {\parbox{\textwidth}{Jason A. S. Hunt$^{1,2}$\thanks{E-mail: j.a.hunt@surrey.ac.uk}, Adrian M. Price-Whelan$^2$, Kathryn V. Johnston$^{3,2}$,\\ Rachel L. McClure$^4$, Carrie Filion$^5$, Ben Cassese$^3$ \& Danny Horta$^2$}\vspace{0.5cm}
\\
$^1$ School of Mathematics \& Physics, University of Surrey, Guildford, GU2 7XH, UK\\
$^{2}$ Center for Computational Astrophysics, Flatiron Institute, 162 5th Av., New York City, NY 10010, USA\\
$^3$ Department of Astronomy, Columbia University, New York, NY 10027, USA\\
$^4$ Department of Astronomy, University of Wisconsin–Madison, Madison, WI, 53706\\
$^5$ William H. Miller III Department of Physics \& Astronomy, The Johns Hopkins University, Baltimore, MD 21218\\
}
\date{Submitted to MNRAS Sept. 6$^{th}$ 2023, accepted Dec. 18$^{th}$ 2023} 

\pagerange{\pageref{firstpage}--\pageref{lastpage}}
\pubyear{2023}

\begin{document}

\maketitle

\label{firstpage}

\begin{abstract}
The second data release of ESA's $Gaia$ mission revealed numerous signatures of disequilibrium in the Milky Way's disc.
These signatures are seen in the planar kinematics of stars, which manifest as ridges and ripples in $R-v_{\phi}$, and in vertical kinematics, where a prominent spiral is seen in the \zvz\ phase space. In this work, we show an equivalent \rvr\ phase spiral forms following a perturbation to the disc. We demonstrate the behaviour of the \rvr\ phase spirals in both a toy model and a high resolution $N$-body simulation of a satellite interaction. We then confront these models with the data, where we find partial \rvr\ phase spirals in the Solar neighborhood using the most recent data from $Gaia$ DR3. This structure indicates ongoing radial phase mixing in the Galactic disc, suggesting a history of recent perturbations, either through internal or external (e.g., satellite) processes.
Future work modelling the \zvz\ and \rvr\ phase spirals in tandem may help break degeneracy's between possible origins of the perturbation.
\end{abstract}


\section{Introduction}
The European Space Agency's $Gaia$ mission \citep{GaiaMission} has revolutionised studies of stellar dynamics in the Milky Way by measuring the position and motion of $\sim1.8\times10^9$ stars in our Galaxy. In turn, this detailed catalogue has revealed the degree of disequilibrium present in the Galactic disc, motivating a need to develop modelling and analysis techniques capable of quantifying and interpreting disequilibrium dynamics. 

One of the most striking disequilibrium features is the \zvz\ phase spiral discovered by \cite{Antoja+18} in $Gaia$ DR2 \citep{DR2}, which showed that the local Milky Way disc is phase mixing vertically following some perturbation. Later studies have shown such phase spirals are present over several kpc in the Galactic disc \citep[e.g.][]{Hunt+22,Antoja23,Frankel23,Alinder+23}, and simulations have shown that we expect them to occur across the entire disc \citep[e.g.][]{BlandHawthorn2021,Hunt+21}. There are currently many proposed explanations for the origins of the \zvz\ phase spirals: The Sagittarius dwarf galaxy, \citep[e.g.][]{Antoja+18,Laporte+18a,Hunt+21}, bar buckling \citep{Khoperskov+19-buclking}, torque from a triaxial halo \citep{Grand+23}, stochastic perturbation from a dark matter subhalo population \citep{Tremaine+23}, or transient spiral structure \citep{Hunt+22}.
    
Regardless of whether the \zvz\ phase spirals are instigated by one of the above processes or a complex combination of them \citep[e.g.][]{Garcia-Conde+23}, the same perturbing influence is likely to impart a force to the radial and azimuthal components of a star's orbit. There has long been evidence that the Milky Way is `ringing' following some perturbation \citep{Minchev09}, and $Gaia$ has now revealed large scale departures from equilibrium in radial and azimuthal stellar kinematics such as the ridges in $R-v_{\phi}$ \citep[e.g.][]{Antoja+18,KBCCGHS18}, which can also be created by the Galactic bar \citep[e.g.][]{Fragkoudi+19}, spiral arms \citep[e.g.][]{Hunt+18,Hunt+19} or interaction with a satellite \citep[e.g.][]{Laporte+18a,Khanna+19,Hunt+21}.

What has currently not been explored is whether a phase spiral also exists in the radial position and motion of stars in the disc following some perturbation, analogous to the \zvz\ spiral in the vertical motion. \cite{Belokurov+23} recently discovered such radial phase mixing is ongoing in the accreted part of the stellar halo. They find phase mixing `chevrons' in $r-v_{\mathrm{r}}$ which are likely tidal debris of the ancient merger event known as $Gaia$-Sausage/Enceladus \citep{Belokurov+18,Helmi+18}. While \cite{Belokurov+23} work in spherical coordinates, $r$ and $v_r$, as is appropriate for the stellar halo, similar structures should also be present in the cylindrical radius and radial motion $R$ and $v_{\mathrm{R}}$ of stars in the disc. We would expect a radial phase spiral to be a natural consequence of radial phase mixing following some more recent perturbations to the in-situ disc stars, e.g. from the Sagittarius dwarf galaxy, or the Milky Way's bar and spiral arms.

The $R-v_{\mathrm{R}}$ plane for disc stars has been explored previously \citep[e.g.][who find `cone like' structure as an indication of phase mixing]{Khanna+19}, but owing to high dust extinction in the Galactic mid-plane, it is challenging to resolve such structure further than $\sim2$ kpc, and the distribution is dominated by stars close to the Sun. However, the equivalent plane to \zvz\ is not $R-v_{\mathrm{R}}$, but rather $\Delta R-v_{\mathrm{R}}$, where $\Delta R=R-R_{\mathrm{G}}$, e.g. capturing the amplitude and phase of the radial epicyclic oscillation around the Guiding radius, $R_{\mathrm{G}}$. \rvr\ has the advantage that stars near the Sun have a range of radial eccentricities and phases, and we can probe a range of $\Delta R$ even with a local sample.

In this work we reveal \rvr\ phase spirals in simulations and in the Galactic disc in $Gaia$ DR3 \citep{DR3}. In Section \ref{orbits} we illustrate the shape of orbits in the \rvr\ plane compared to the \zvz\ plane. In Section \ref{sim} we describe a toy model (\ref{toy-model}) and the $N$-body simulation used in this paper (\ref{nbody}), and in Section \ref{rvrsim} we show the spirals and substructure in the \rvr\ plane both locally and across the disc of the simulation. In Section \ref{data} we show the \rvr\ plane data from $Gaia$ DR3. We describe our treatment of the $Gaia$ data in Section \ref{data-selection}, and in Section \ref{rvrdata} we show the \rvr\ plane in the Solar neighborhood and over 2 kpc. In Section \ref{dlz} we decompose the 2 kpc sample as a function of angular momentum, and the describe the introduced biases in Section \ref{bias}. In Section \ref{utility} we discuss what we can learn from this projection, and in Section \ref{conc} we summarise our conclusions.

\section{Orbits and epicycles in an axisymmetric potential}\label{orbits}
In this section we introduce the $\Delta R - v_{\mathrm{R}}$ phase plane and illustrate how individual disc stars evolve along their orbits in the absence of any external perturbations.

Stars on sufficiently circular orbits move along nearly closed curves in the $z - v_z$ plane as they oscillate about the Galactic midplane. Similarly, stars evolve along nearly closed curves in $\Delta R - v_{\mathrm{R}}$ as they progress along their radial epicycles centered on their guiding radius, $R_{\mathrm{G}}$. In both cases, as the orbits become less circular (higher action), these trajectories in phase space become less like simple harmonic motion ellipses and distort into more complex shapes. 

\begin{figure}
\centering
\includegraphics[width=\hsize]{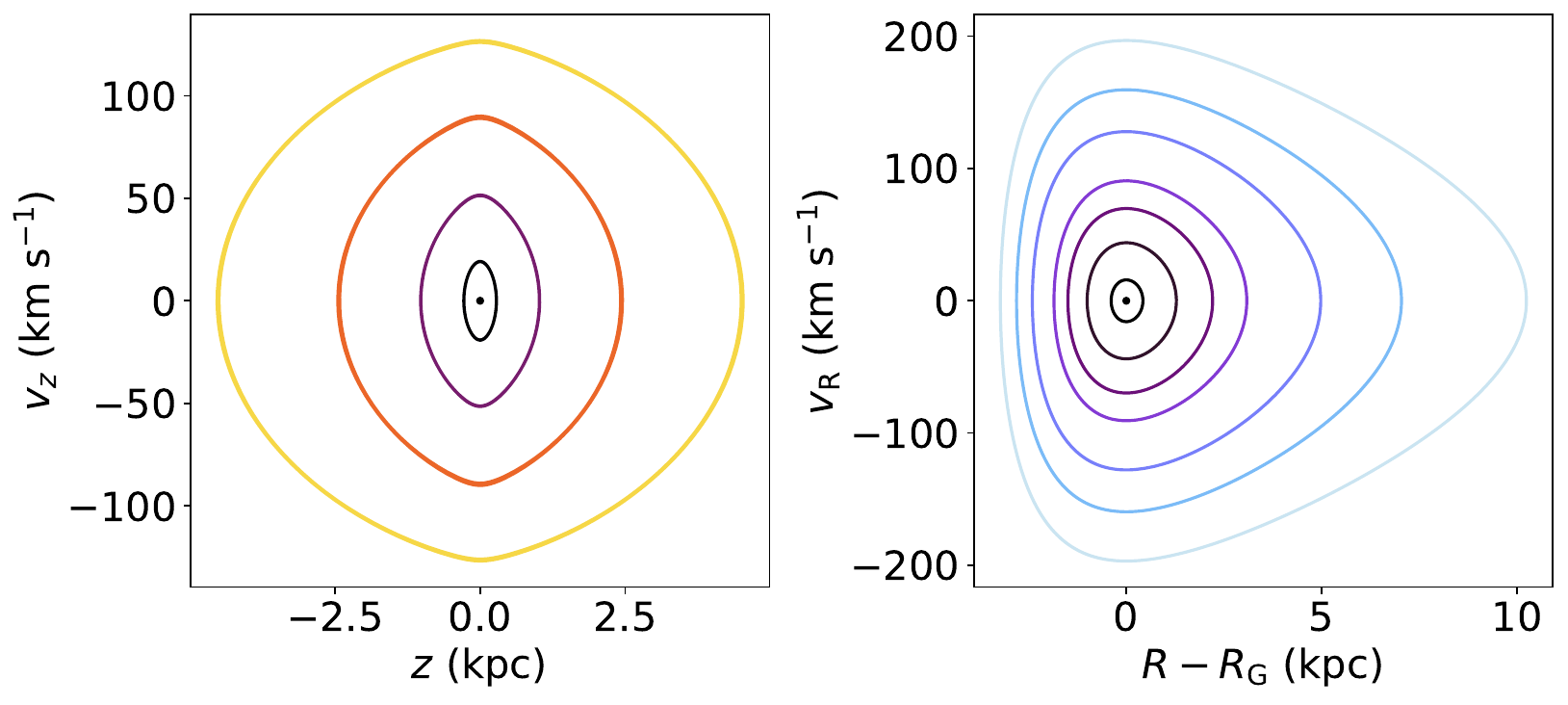}
\caption{\textbf{Left:} Four example orbits, each with the same $J_{\phi}$ and $J_{\mathrm{R}}=0$ and increasing $J_z$ (dark to light). \textbf{Right:} Six orbits with identical $J_{\phi}$ and $J_z = 0$, but increasing $J_{\mathrm{R}}$ (dark to light). Note that in both cases for orbits with higher action, the trajectories remain nearly closed but increasingly diverge from ellipses in a different way.}
\label{orbit_trajectories}
\end{figure}

Figure \ref{orbit_trajectories} shows example orbits calculated in the \texttt{MilkyWayPotential2022}\ potential from \texttt{gala}\ \citep{gala}. Each orbit has the same $J_{\phi}$=1900 km s$^{-1}$ kpc$^{-1}$. The left panel shows orbits with zero radial action, $J_{\mathrm{R}}=0$, and increasing vertical action (dark to light) in the \zvz\ plane. The right panel shows orbits with zero vertical action, $J_{\mathrm{z}}=0$, and increasing radial action (dark to light) in the \rvr\ plane.

Note the difference in the shape of the distortion for orbits in the \rvr\ and \zvz\ planes. Since the effective potential is symmetric about the midplane, the \zvz\ trajectories remain symmetric even as they diverge from simple ellipses. However, since the effective potential decreases with increasing galactic radius, it is not symmetric about $R_{\mathrm{G}}$ and consequently the expected trajectories are themselves not symmetric. Instead, stars in the \rvr\ plane move along ``arrowhead"-like trajectories, reaching a larger $\Delta R$ when further from the galactic center than when nearer it.


Regardless, in both cases, when considering a phase mixed system of many stars, these shapes can be traced by any label invariant to orbital phase, such as the local stellar density (assuming no selection effects) or stellar abundances \citep{Price-Whelan+21,Horta+23}.

\section{Simulated data}\label{sim}
In this section we explore the creation of \rvr\ phase spirals in a toy model, and a high resolution $N$-body simulation to aid with later comparison and interpretation of the $Gaia$ data.

\subsection{Toy model}\label{toy-model}
As an initial illustration we present a toy model of particles evolved in a fixed Milky Way-like potential following an artificial kick. 

We use \texttt{Agama}\ \citep{agama} to sample $2\times10^7$ particles within a limited $J_\phi$ range of $1760<J_{\phi}<1955$ kpc km s$^{-1}$ from a quasi-isothermal distribution function \citep{B12-1,SB15b-DFs} given by:
\begin{eqnarray}
    f(J_{\mathrm{R}},J_{\phi},J_z) = \frac{1}{8\pi^3}[1+\mathrm{tanh}(J_{\phi}/L_0)]\frac{\Omega}{R^2_{\mathrm{d}}\kappa^2}\mathrm{e}^{-R_{\mathrm{c}}/R_{\mathrm{d}}}\nonumber \\
    \times \frac{\kappa}{\sigma^2_{\mathrm{R}}}\mathrm{e}^{-\kappa J_{\mathrm{R}}/\sigma_{\mathrm{R}}^2}\frac{\nu}{\sigma^2_z}\mathrm{e}^{-\nu J_z/\sigma_z^2},
\end{eqnarray}
where we set $R_{\mathrm{d}}=3.45$ kpc, $R_{\sigma}=7.8$ kpc, $\sigma_{\mathrm{R},0}=48.3$ km s$^{-1}$, $\sigma_{z,0}=30.7$ km s$^{-1}$ following the values for the thin disc in Table 3 of \cite{SB15b-DFs}. The values for $\kappa$, $\nu$ and $R_{\mathrm{c}}$ are calculated from the \texttt{MilkyWayPotential2022}\ potential from \texttt{gala}\ \citep{gala}, as described in \cite{Hunt+22}.

\subsubsection{Phase mixing following a kick}
We apply three different impulsive `kicks' by offsetting the radial velocities from the equilibrium distribution by 25, 50 and 75 km s$^{-1}$, and then integrate the sample within the \texttt{MilkyWayPotential2022}\ for 1 Gyr.

\begin{figure}
\centering
\includegraphics[width=\hsize]{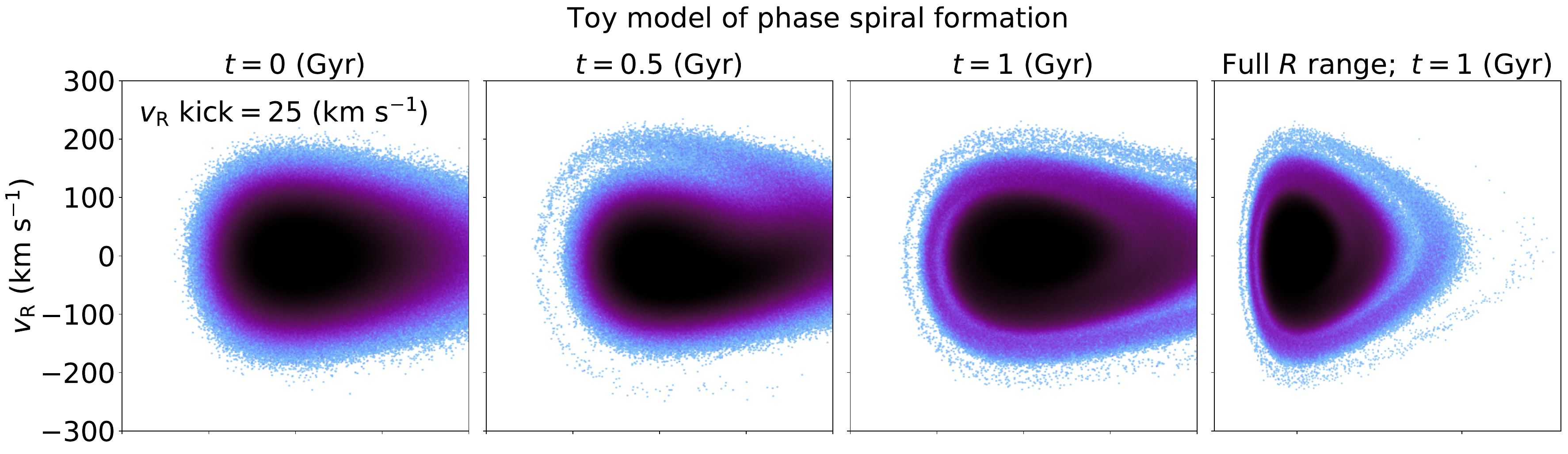}
\includegraphics[width=\hsize]{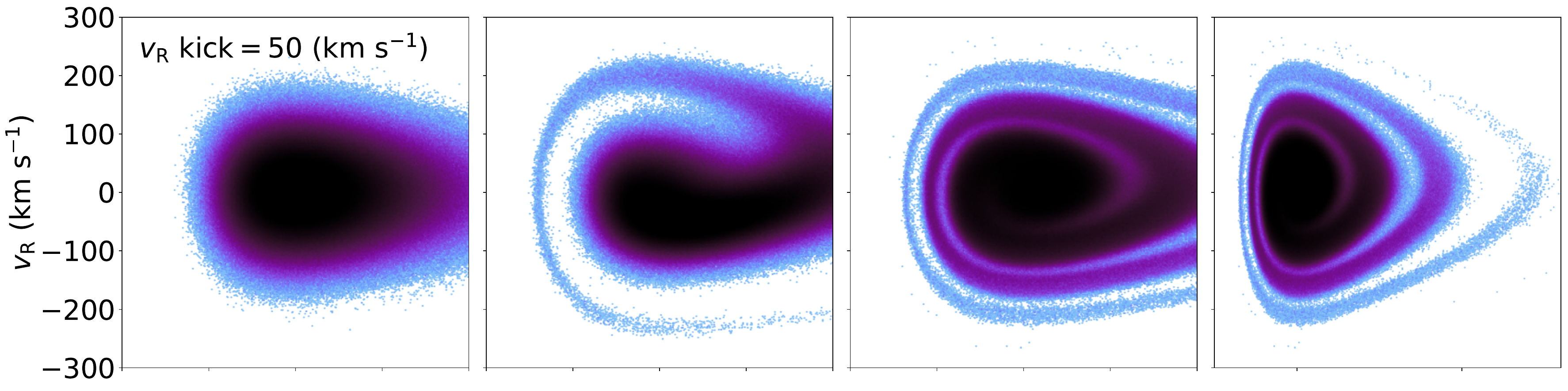}
\includegraphics[width=\hsize]{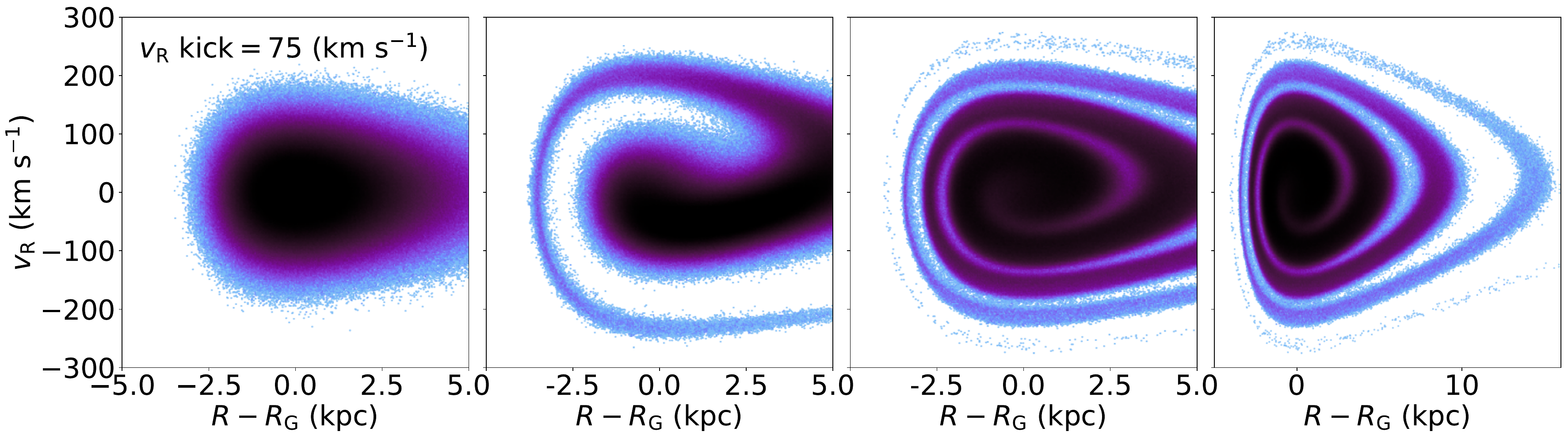}
\includegraphics[width=\hsize]{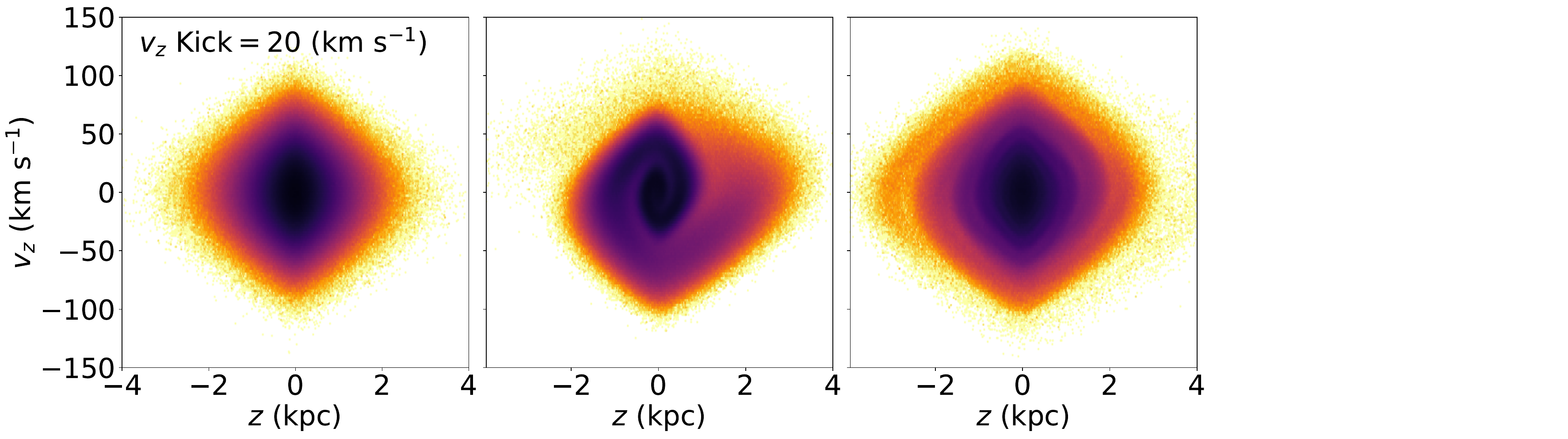}
\caption{Toy model of $2\times10^7$ particles in a fixed potential showing \rvr\ phase spirals that form following an impulsive kick of 25 (top row), 50 (second row) and 75 (third row) km s$^{-1}$. The left column shows the initial equilibrium distribution, the second and third columns show the \rvr\ phase spiral after 0.5 and 1 Gyr respectively, over the same range as Figure \ref{data-split} for later comparison. The rightmost column shows the full radial extent of the \rvr\ spiral after 1 Gyr. The bottom row shows the same sample kicked vertically by 20 km s$^{-1}$, for comparison.}
\label{toy}
\end{figure}

Figure \ref{toy} shows the initial distribution of the sample in \rvr\ (left column), the sample after 500 Myr (second column) and after 1 Gyr (third column), for the sample with the 25 km s$^{-1}$ (upper row), 50 km s$^{-1}$ (second row), and 75 km s$^{-1}$ (third row) kicks. The range of the left three columns is chosen for later comparison with the data, and the right hand column shows the full distribution for completeness. The lower row of Figure \ref{toy} shows the same sample kicked by 20 km s$^{-1}$ in the vertical direction, for comparison.

Figure \ref{toy} shows that phase spirals do form in \rvr\ as the samples phase mix back towards equilibrium following the `kick', as expected. The trend of increasingly strong \rvr\ phase spirals with increasingly strong kicks is clear, and unsurprising. While the amplitude of the spiral changes, the general morphology does not, as it depends on the shape of the galactic potential and the direction of the `kick', which remain fixed between samples. The \rvr\ phase spirals in the stronger cases are quite striking, but this toy example shows only a limited sample without selection effects, and without self-gravity, which is known to be important in phase spiral formation and morphology \citep{Widrow23}.

Note that the distribution in \rvr\ is asymmetric in the radial dimension in both the initial sample, (which matches the shape of the orbits in the right panel of Figure \ref{orbit_trajectories}) and following the perturbation, owing to the radially decreasing nature of the galactic potential. This is in contrast to the symmetric nature of the initial \zvz\ plane shown in the lower row (although the kick and subsequent phase mixing makes the \zvz\ distribution asymmetric until it phase mixes), owing to the symmetric nature of the vertical potential. This illustrates that while phase mixing and thus phase spirals occur both radially and vertically, we do not expect them to produce phase spirals with the same morphology. 

Also, in this simple toy model a vertical kick has almost no discernible effect on the morphology of the \rvr\ spiral, and similarly a radial kick has almost no discernible effect on the morphology of the \zvz\ spiral. In reality we would expect a perturber such as a satellite galaxy to impart a radial and vertical kick together, and we would expect there to be some coupling in the response in a self-gravitating disc. However, this simple toy is enough for now to illustrate the expected shape and properties of a $R-v_{\mathrm{R}}$ phase spiral, and we defer a more thorough exploration of coupling to a future work.

\subsection{The $N$-body simulation}\label{nbody}
To explore \rvr\ phase spiral formation in a more realistic model, we make use of the $N$-body simulation M1 from \cite{Hunt+21}, which consists of a merger of a dwarf galaxy into a highly stable disc. Full details of the simulation setup and evolution are available in \cite{Hunt+21}, but in brief; the host galaxy is based on the MWb model from \cite{WD05} with total mass $\sim6\times10^{11}$ $M_{\odot}$, consisting of a $8.8\times10^8$ particle live NFW halo \citep{NFW97}, a $2.2\times10^7$ particle Hernquist bulge \citep{H90} and a $2.2\times10^8$ particle exponential disc with a high Toomre parameter $Q=\sigma_{\mathrm{R}}\kappa/3.36G\Sigma=2.2$ \citep{Toomre64} which is stable against bar and spiral formation for several Gyr without the addition of the satellite perturber \citep{Hunt+21}. The satellite is the L2 model from \cite{LJGG-CB18}, which is comprised of two Hernquist spheres \citep{H90}, with a virial mass of $\sim6\times10^{10}$ $M_{\odot}$, making the merger approximately 1-10 mass ratio.

The combined model is evolved with the $N$-body tree code \texttt{Bonsai} \citep{Bonsai,Bonsai-242bil} for 8.292 Gyr, during which period the satellite merges into the host galaxy. In this instance we choose to work with both an `early time' snapshot 454 at $t=4.45$ Gyr, which has experienced a single pericentric passage of the satellite 2 Gyr previously, and a `late time' snapshot 720, at $t=7.05$ Gyr owing to the presence of both a tightly wound \zvz\ phase spiral such as seen in the $Gaia$ data, and the presence of a \rvr\ phase spiral in the `Solar neighborhood' when assuming the observer to be located at $(x,y,z)_{\odot}=(-8.2,0,0.02)$ kpc. In the `late time' snapshot the dwarf galaxy is located at $(x,y,z,v_x,v_y,v_z)=(-8.1,1.2,24.8,-117.4,8.6,-30.2)$ kpc, km s$^{-1}$ compared to the estimated position and motion of the Sagittarius remnant $(x,y,z,v_x,v_y,v_z)=(9.5,-0.4,-6.7,115.7,5.9,311.9)$ kpc, km s$^{-1}$ \citep{Vasiliev+20}. Thus, the dwarf galaxy in the Simulation is not at the same phase as Sagittarius, and we are not attempting to reproduce the conditions in the Milky Way, merely illustrate the phase spirals following multiple perturbations.

In general, we stress that this model is not designed to specifically reproduce the Milky Way Sagittarius interaction \citep[for which we refer the reader to the models presented in][]{BBH21,Stelea23}, but rather be a general laboratory for investigating merger dynamics in an otherwise quiet disc. The satellite in model M1 has experienced seven disc crossings by the `late time' snapshot and is about to cross again in $\sim 150$ Myr \citep[see][for the satellite orbit and mass loss history]{Hunt+21}. While the exact number of disc crossing experienced by the Sagittarius dwarf is not fully known \citep[although it can be constrained by changes in star formation and disk velocity dispersion; e.g.][]{RuizLara+20,Das+23} it has certainly experienced multiple passages since first infall.


To calculate action-angle variables for the simulation we first use \texttt{Agama}\ \citep{agama} to reconstruct the simulation potential as a composite of two multipole expansions for the stellar bulge and dark matter halo, and an axisymmetric \texttt{CylSpline}\ expansion for the disc. We then use \texttt{Agama}'s \texttt{ActionFinder}\ routine to calculate actions, angles and frequencies in the reconstructed potential.

The full time evolution of simulation M1 from \cite{Hunt+21} in 10 Myr snapshots is now freely available at \texttt{SciServer}\footnote{https://sciserver.org/datasets/smudge/} \citep{sciserver}, along with the rest of the `SMUDGE' (Satellite Mergers Usher Disc Galaxy Evolution) simulation suite M2, D1 \& D2. The simulations can be accessed and analysed within the \texttt{SciServer} Compute app and the package for interfacing with the simulations is introduced and explained in a provided example \texttt{Jupyter Notebook}, available both on \texttt{SciServer} and \texttt{GitHub}\footnote{https://github.com/CarrieFilion/SMUDGE}.

\begin{figure}
\centering
\includegraphics[width=\hsize]{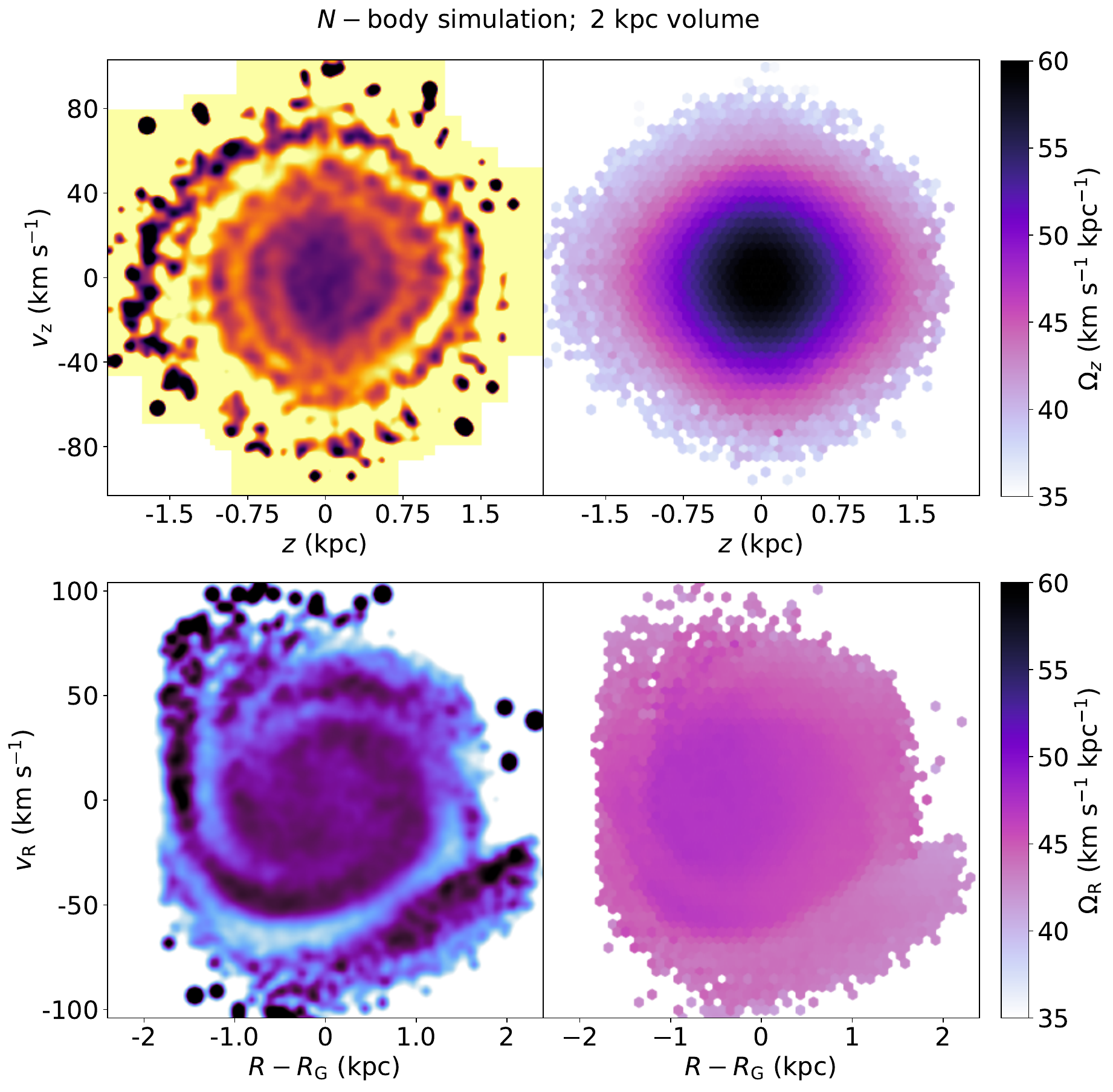}
\caption{\textbf{Upper row:} $z-v_z$ phase spiral selected within a local volume in the `late time' simulation snapshot, shown in density (left) and as a function of $\Omega_z$ (right). \textbf{Lower row:} \rvr\ phase spiral within the same volume in the simulation, as a function of density (left) and $\Omega_{\mathrm{R}}$ (right). The \zvz\ phase spiral winds up more rapidly than the \rvr\ spiral as the vertical frequency gradient with respect to $J_z$ is much steeper than the radial frequency gradient with respect to $J_{\mathrm{R}}$, as illustrated in the right column, where $\Omega_z$ and $\Omega_{\mathrm{R}}$ are shown on the same scale.}
\label{model-local}
\end{figure}

\begin{figure}
\centering
\includegraphics[width=\hsize]{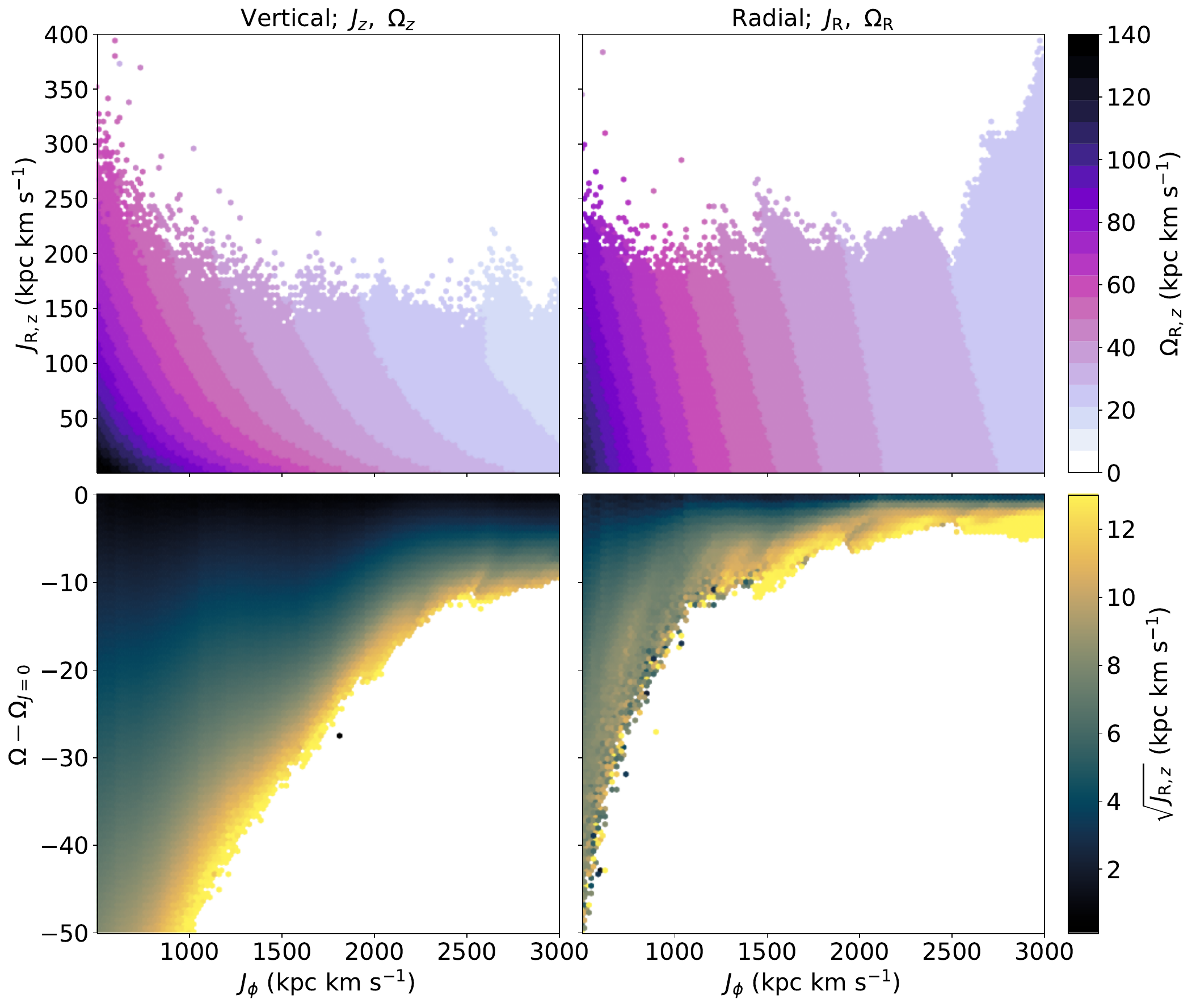}
\caption{Frequency and action gradients for disc stars in the simulation M1 \citep{Hunt+21}. \textbf{Upper row:} Vertical (left) and radial (right) action as a function of angular momentum, $J_{\phi}$, colored by the vertical and radial frequency, $\Omega_z$, $\Omega_{\mathrm{R}}$ respectively. \textbf{Lower row:} Vertical (left) and radial (right) frequency subtracting the asymptotic vertical and radial frequency at $J=0$, colored by $\sqrt{J_z}$ and $\sqrt{J_{\mathrm{R}}}$ respectively.}
\label{freqs}
\end{figure}

\subsubsection{\rvr\ structures in the simulation}\label{rvrsim}
In \cite{Hunt+21} we presented an analysis of the \zvz\ phase spiral across the disc of the simulation M1. As discussed in the introduction and shown above in the toy model, such features should also appear in $\Delta R-v_{\mathrm{R}}$ (the analogous space to $z-v_z$), where $\Delta R=R-R_{\mathrm{G}}$, describing the stars oscillation around their guiding radius.

As an initial illustration, Figure \ref{model-local} shows the well known $z-v_z$ plane (upper left) and the equivalent \rvr\ plane (lower left) for disc stars in the `late time' snapshot within a local volume of 2 kpc selected around $(x,y,z)=(7.2,0,0)$ kpc, and with Guiding radii $7<R_{\mathrm{G}}<7.4$, i.e. within 0.2 kpc of the `local' radius. This is an arbitrary choice in the simulation, chosen merely as an illustration of a volume which contains two clear phase spirals. 

For visualisation purposes we have subtracted a Gaussian smoothed background distribution, such that the left column shows the difference in density between the data and the smoothed version. The smoothing, is different in \zvz\ (20\% of axis range) and \rvr\ (30\% of axis range), although all subsequent \rvr\ panels have the same smoothing. The $z-v_z$ phase spiral in the top left panel is qualitatively similar to the local $Gaia$ data, it has been reproduced in many simulations and has been discussed in numerous other works \citep[e.g.][]{Antoja+18,Laporte+18a,Hunt+21,BlandHawthorn2021}.

\begin{figure*}
\centering
\includegraphics[width=\hsize]{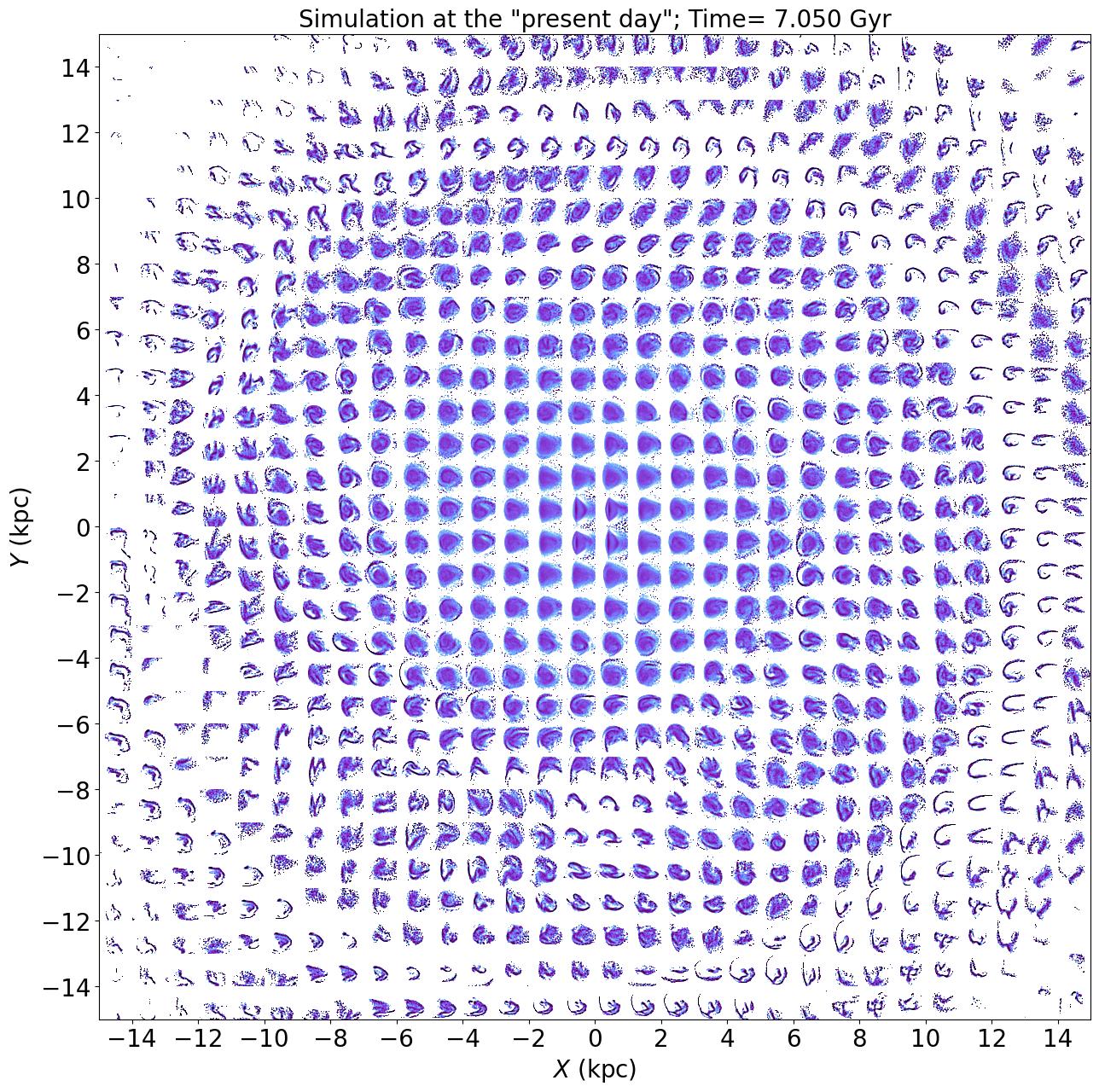}
\caption{$\Delta R-v_{\mathrm{R}}$ planes for 900 (30 by 30) 1 kpc-square regions in the disc at snapshot 720 ($t=7.05$ Gyr) from simulation M1 from \citet{Hunt+21}. 
The colormap in each subpanel corresponds to the number of particles in each bin of the \rvr\ phase space, with darker color corresponding to more particles.
The structure of the perturbed phase-space distribution is more irregular at larger radius because the self-gravity of the disc is weaker.
The time evolution of \rvr\ and \zvz\ can be seen side by side in M1-vertical-radial at \url{https://zenodo.org/records/8402668}.}
\label{rvr-large720}
\end{figure*}

\begin{figure*}
\centering
\includegraphics[width=\hsize]{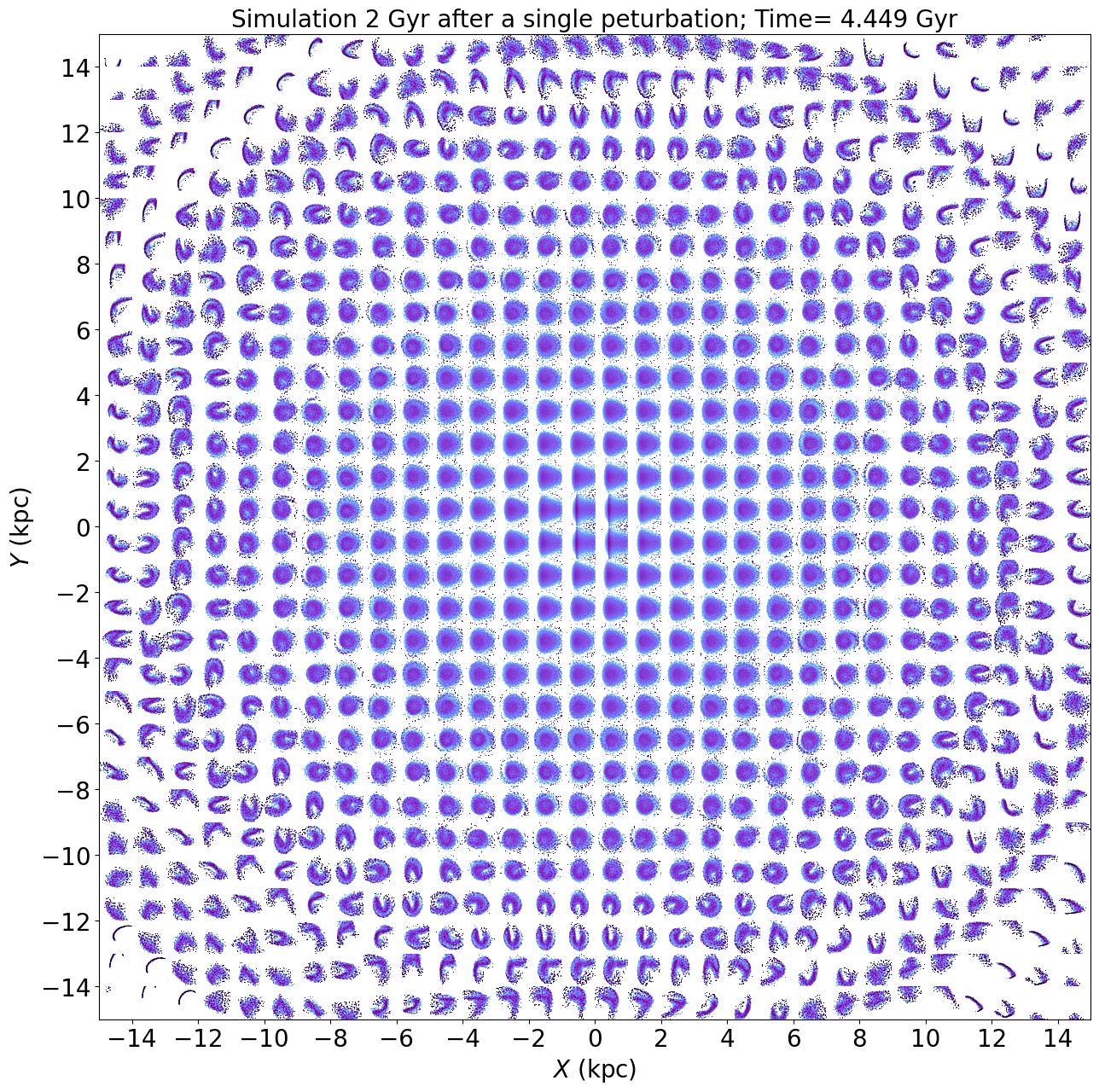}
\caption{Same as Figure \ref{rvr-large720} but occurring earlier in the simulation at time 4.45 Gyr, 2 Gyr after the first passage of the satellite galaxy, but before the second passage \citep[see][for the orbit of the satellite]{Hunt+21}.}
\label{rvr-large454}
\end{figure*}

Any perturbation to the disc should phase mix away vertically, azimuthally and radially, and thus such a \rvr\ phase spiral should be expected, as shown in the toy model in Section \ref{toy-model}. However, the $z-v_z$ phase spiral is especially striking because the vertical frequency of stars, $\Omega_z$, is a strong function of vertical action, $J_z$, as shown in the upper right panel of Figure \ref{model-local} (where $J_z$ is approximately the distance from the centre of the panel). This causes a perturbation to rapidly `wind up' into a spiral pattern. The radial frequency of stars, $\Omega_{\mathrm{R}}$ is a much weaker function of the radial action, $J_{\mathrm{R}}$, as shown in the lower right panel of Figure \ref{model-local}.  

To further illustrate this, the upper row of Figure \ref{freqs} shows the vertical (left) and radial (right) action as a function of angular momentum, $J_{\phi}$, colored by the vertical and radial frequencies $\Omega_z$ and $\Omega_{\mathrm{R}}$, respectively, for disc stars in the simulation M1. The upper left panel shows that $\Omega_z$ is a strong function of both $J_{\phi}$ and $J_z$. In contrast, the upper right panel shows that while $\Omega_{\mathrm{R}}$ is also a strong function of $J_{\phi}$, the gradient of $\Omega_{\mathrm{R}}$ with respect to $J_{\mathrm{R}}$ is much shallower. However, while shallow, there is a gradient, which is required for phase spiral formation.

The lower row of Figure \ref{freqs} shows the difference between the vertical (left) and radial (right) frequencies of the disc stars and the asymptotic frequency, $\nu=\Omega_z|J_z=0$ (km s$^{-1}$ kpc$^{-1}$) and $\kappa=\Omega_{\mathrm{R}}|J_{\mathrm{R}}=0$ (km s$^{-1}$ kpc$^{-1}$), respectively, for a given $J_{\phi}$. The lower left panel shows that the spread between the asymptotic vertical frequency, $\nu$, as a function of action is much larger than the spread between the asymptotic radial frequency, $\kappa$, at a given $J_{\phi}$. 

Thus, the timescale for mixing in this space is much longer than the vertical phase mixing in the $z-v_z$ plane, resulting in a much more open pitch angle of the spiral in \rvr\ compared to \zvz (as also shown when comparing the toy models for \rvr\ and \zvz\ in Figure \ref{toy}). Given the long time scale we would expect the presence of bar resonances and spiral arms to readily interfere with older patterns, but perhaps we can still exploit any structure in \rvr\ much as we have done with the phase spirals in \zvz.

In order to test whether such \rvr\ phase spirals are common within the simulation we relax the local selection. Following \cite{Hunt+21}, Figure \ref{rvr-large720} shows the \rvr\ distribution in 30x30 1 kpc bins across the face of the disc for the late time snapshot at 7.05 Gyr. Figure \ref{rvr-large720} shows a large variation in the morphology of the individual \rvr\ planes. Some show spiral-like patterns (most commonly around a radius of $\sim7$ kpc), while others are chaotic, or simply arcs (more commonly in the outer disc around $\sim13$ kpc). So, `clean' \rvr\ phase spirals such as are shown in Figure \ref{model-local} are not `common' within the simulation snapshot, but neither are they particularly rare, or unique. When examining the time evolution of the local $R-v_{\mathrm{R}}$ plane$^3$, the Solar neighborhood does not always contain \rvr\ phase spirals. Some snapshots contain complex structure in the \rvr\ plane but no phase spirals, as we would expect if they are disrupted by secular dynamics such as the bar resonances or the passage of a spiral arm.

In \cite{Hunt+21} we showed that the \zvz\ phase spirals became increasingly messy following repeated satellite impacts, while earlier snapshots showed well defined spirals after the first impact alone \citep[see Figures 7 \& 9 in][]{Hunt+21}. As such, Figure \ref{rvr-large454} shows the same plane of \rvr\ distributions at an earlier time ($t=4.49$ Gyr). This snapshot is 2 Gyr after the first impact and shortly before the second interaction. Around the `Solar radius' ($R\sim8$ kpc) many weak partial phase spirals have formed, while the outer disc is dominated by arcs of varying completeness, resembling ``pac-men", or ``fortune cookies", instead of spirals. As expected, the shallow gradient of $\Omega_{\mathrm{R}}$ as a function of $J_{\mathrm{R}}$ (shown in Figure \ref{model-local}) has resulted in only partial phase mixing (and minimal spiral formation) especially in the outer galaxy where stellar frequencies are lower, even over 2 Gyr.

While the \rvr\ planes do not show `clean' phase spirals such as are seen in the \zvz\ direction \citep[see Figures 7 and 9 of][or the animated evolution of both side by side here\footnote{See M1-vertical-radial at \url{https://zenodo.org/records/8402668}}]{Hunt+21}, it remains clear that the passage of the satellite galaxy imparts coherent substructure in the radial oscillation of stars as well as the vertical, as expected. Note that the individual \rvr\ planes trace out galaxy spanning spiral patterns, similar to the \zvz\ planes shown in \cite{Hunt+21}. 

Most importantly, these \rvr\ features can perhaps be exploited in a similar fashion as is being done in studies of the \zvz\ spirals to infer information about the Galactic potential, or the nature of the perturbation \citep[e.g.][]{WidmarkIV,Frankel23,Darragh-Ford+23,Guo+23} if such structures are also seen in the data. 

\section{$Gaia$ DR3 data}\label{data}
In this section we describe our treatment of the \textit{Gaia} DR3 data, and the \rvr\ substructure in the Solar neighborhood and beyond.

\subsection{Data selection and treatment}\label{data-selection}
We select stars from the \textit{Gaia} DR3 catalogue that have radial velocities with radial velocity error, $\sigma_{v_{\mathrm{R}}}<5$ km s$^{-1}$, a fractional parallax error ($\sigma_\varpi$) of less than 10 per cent (or $\varpi / \sigma_\varpi > 10$), and the renormalised unit weighted error, $RUWE<1.4$, providing an overall sample of 16,673,097 stars. We calculate distances from simple parallax inversion, which should be reliable for the this nearby sample used in this work. We use \texttt{astropy}\ \citep{astropy:2022} to transform the sample to Galactocentric Cartesian and cylindrical coordinates assuming  a sun--Galactic centre distance of $R_0=8.275~\kpc$ \citep{GRAVITY:2021}, a Solar height above the midplane of $20.8~\pc$ \citep{BB19}, and a total Solar velocity with respect to the Galactic center of $\boldsymbol{v}_\odot = (8.4, 251.8, 8.4)~\kms$ calculated from the apparent radial velocity and proper motion of Sgr A$^*$ \citep[see][for the source data and calculation]{Reid:2020,GRAVITY:2021,Hunt+22}.

We use the ``St\"ackel Fudge'' \citep{Binney:2012, Sanders:2012} as implemented in \texttt{galpy} \citep{galpy} to compute actions, $\boldsymbol{J} = (J_{\mathrm{R}}, J_{\phi}, J_z)$, their conjugate phase angles, $\boldsymbol{\theta} = (\theta_{\mathrm{R}}, \theta_{\phi}, \theta_z)$ and frequencies $\boldsymbol{\Omega}=(\Omega_{\mathrm{R}}, \Omega_{\phi}, \Omega_z)$ using the \texttt{MilkyWayPotential2022}\ from \texttt{gala}. We calculate the guiding radius of stars, $R_{\mathrm{G}}=J_{\phi}/v_{\mathrm{circ}}(\boldsymbol{r})$, where $v_{\mathrm{circ}}(\boldsymbol{r})$ is calculated from \texttt{MilkyWayPotential2022}.

\subsection{\rvr\ structures in the $Gaia$ data}\label{rvrdata}
In this section we show the dynamical structure present in \rvr\ in the Solar neighborhood, and as a function of angular momentum.

\begin{figure}
\centering
\includegraphics[width=\hsize]{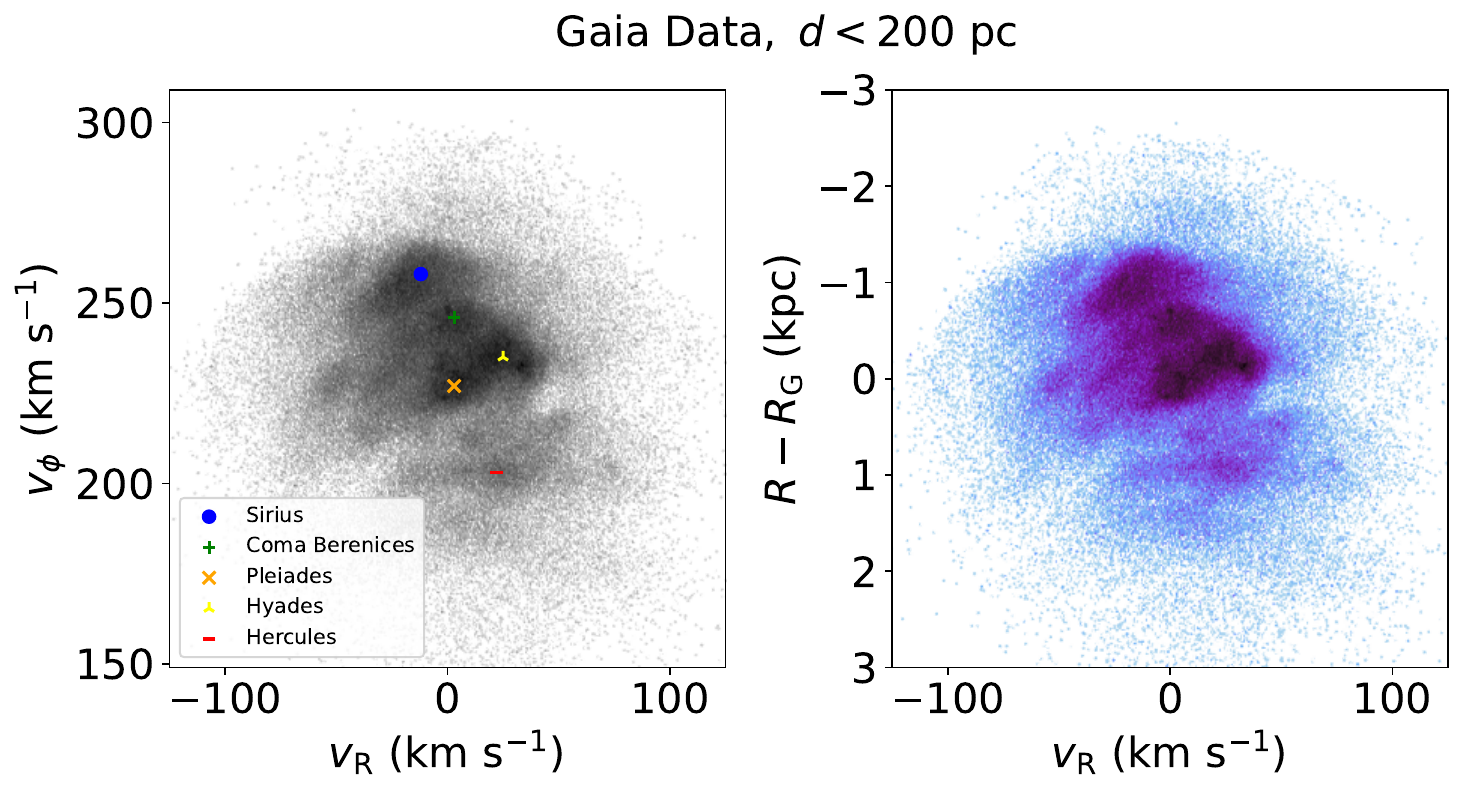}
\caption{$v_{\mathrm{R}}-v_{\phi}$ (left) and $v_{\mathrm{R}}-\Delta R$ (right) for a 200 pc volume around the Sun. Note that over such a local volume, they are almost identical. The classical moving groups are marked in the left panel.}
\label{moving-groups}
\end{figure}

As an initial illustration, the left panel of Figure \ref{moving-groups} shows the local $v_{\mathrm{R}}-v_{\phi}$ plane for all stars within 200 pc. The classical moving groups originally discovered in a series of papers by Olin Eggen \citep[see][and references within for a summary]{Eggen1996} are clearly visible, and have been marked. The right panel shows the \rvr\ plane (axes swapped to match $v_{\mathrm{R}}$) for the same 200 pc sample, in which the structure is clearly the same. This is to be expected within a very local volume, because we see only a small section of these dynamical structures even though we know they span several kpc across the Galactic disc \citep[e.g.][]{KBCCGHS18,AntojaeDR321}. The \rvr\ panel is effectively an orbit map, yet within a very local volume stars on the same orbit must have the same kinematics, and very little is learned beyond the classic $v_{\mathrm{R}}-v_{\phi}$ plane.

For example, at a single radius there is a clear relation between $v_{
\phi}$ and $\Delta R$. Only a single $v_{\phi}$ will lead to a circular orbit ($\Delta R=0$), and a star with higher or lower $v_{\phi}$ will have $\Delta R$ negative or positive respectively, while on the inner or outer part of their radial epicycles. However, when analysing a larger volume, orbit maps become more informative than kinematic maps \citep[see e.g.][]{Hunt+20}. For example, when exploring a large range of $R$, there are many $v_{\phi}$ which can correspond to $\Delta R=0$ for stars at different radii (assuming a non-flat rotation curve), and $v_{\phi}$ is no longer a direct map to $\Delta R$ (for any radial eccentricity). The effect over a small radial range is minimal, but as we expand our sample to increasingly large distances $\Delta R$ will increasingly diverge from $v_{\phi}$.

\begin{figure}
\centering
\includegraphics[width=\hsize]{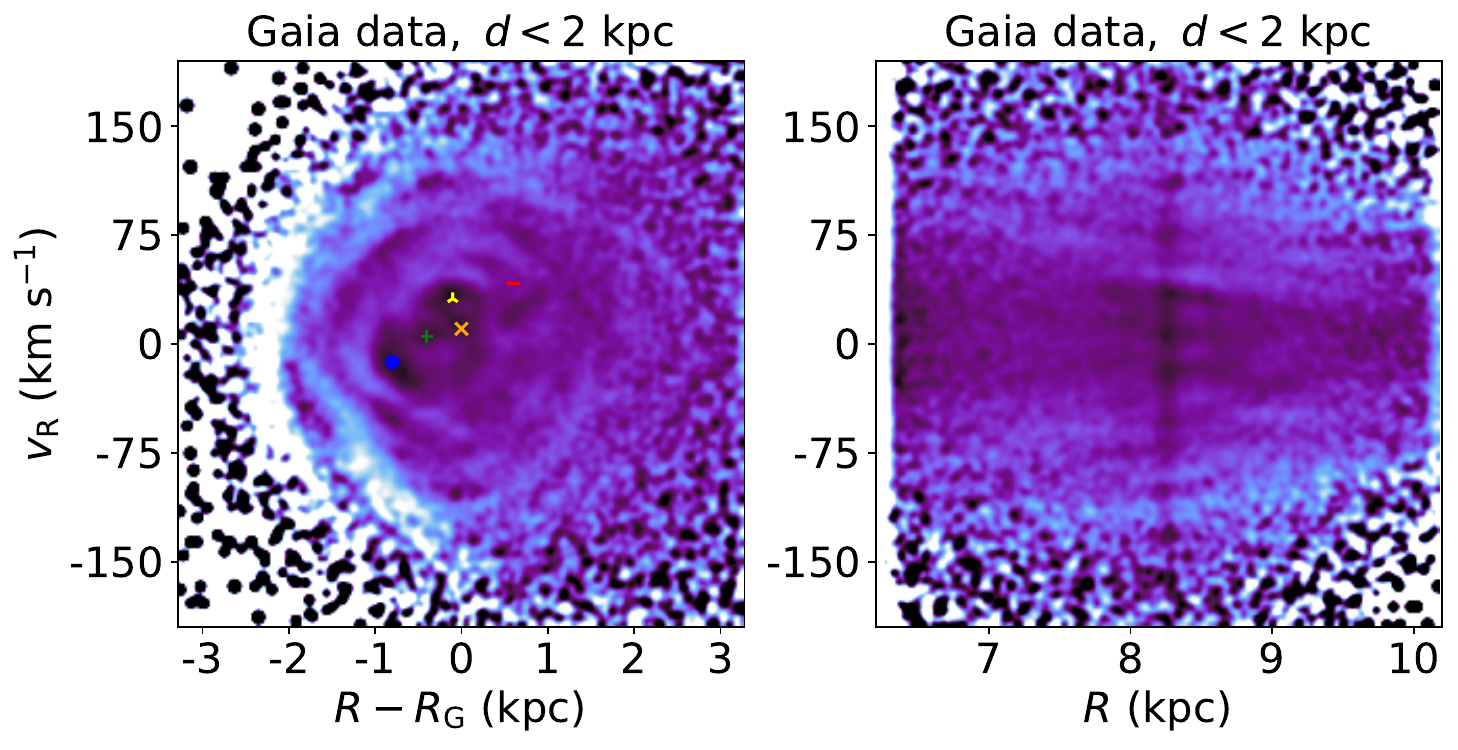}
\caption{\textbf{Left:} \rvr\ plane stars with physical distance $d<2$ kpc from the Sun. As with Figure \ref{model-local}, the colormap shows the residual phase-space density of stars after unsharp-masking the data (i.e. after subtracting a smoothed version of the phase-space density from itself). The classical moving groups, Sirius (blue dot), Coma Berenices (green plus), Pleiades (orange cross), Hyades (yellow tri) and Hercules (red dash) are marked matching Figure \ref{moving-groups}.
\textbf{Right:} $R-v_{\mathrm{R}}$ plane for the same stars, illustrating the difference in $R-v_{\mathrm{R}}$ and \rvr\ for a local volume}
\label{data-local}
\end{figure}

As such, the left panel of Figure \ref{data-local} shows the \rvr\ plane for stars within 2 kpc from the Sun. We now recover arcs with higher $\mid\Delta R\mid$, but still we do not find a clean phase spiral, but more of a `rose-like' structure. In addition, the imprint of the well known local moving groups; Hyades, Pleiades, Hercules, etc. remain visible with comparatively low radial eccentricity in the centre of the panel, and have been marked again for comparison with Figure \ref{moving-groups} \citep[note that many of the unmarked features are also previously discovered kinematic groups; see e.g.][but a detailed census of kinematic substructure is not the purpose of this work]{RAF18}. To illustrate the utility of the \rvr\ plane over the $R-v_{\mathrm{R}}$ plane, the right panel of Figure \ref{data-local} shows the $R-v_{\mathrm{R}}$ plane for the same sample of stars. Here, some of the substructure is faintly visible as chevrons \citep[as shown in $r-v_r$ for the halo in][]{Belokurov+23} but it is not possible to resolve any full wraps within a local sample.

While the left panel is not a perfect phase spiral either, we do also expect the \rvr\ plane to be a superposition of many different \rvr\ structures as a function of angular momentum, $L_z$ (or $R_{\mathrm{G}}$), and $\theta_{\phi}$ as has been shown in several works for the \zvz\ phase spiral \citep{Li20,Hunt+21,Gandhi+20,Darragh-Ford+23}.

\subsubsection{As a function of angular momentum, $J_{\phi}$}\label{dlz}
In order to test how the structures vary with angular momentum, Figure \ref{data-split} shows the \rvr\ plane for stars with $d<2$ kpc, in nine guiding radius bins from $6.2<R_{\mathrm{G}}<10.2$ kpc, each selected such that $\mid R_{\mathrm{G}}-R_\mathrm{C}\mid<0.5$ kpc (where $R_\mathrm{C}$ is the centre of the bin). We recover several fragments of arcs and ridges, but no complete phase spirals, owing to the selection function of the sample.

From these partial fragments in phase and angular momentum it's difficult to conclusively state whether the complete \rvr\ planes would form phase spirals, or whether they instead consist solely of arcs and ridges of stars following the same orbital tracks around their radial epicycle. Both cases are present in the late time simulation snapshot presented in Figure \ref{rvr-large720}, which contains many \rvr\ planes that are qualitatively similar to the structure in Figure \ref{data-split} (although the features are finer in the data). Conversely, the early time simulation snapshot in Figure \ref{rvr-large454} predominantly presents single arcs of varying lengths in the outer disc or faint partially wound spirals in the inner disc, neither of which are a good match to the data. 

This is as unsurprising, because the early time snapshot in Figure \ref{rvr-large454} has only experienced one single satellite impact approximately 2 Gyr previously, and the `hotness' of the disc has prevented additional secular perturbations such as from a bar or internally excited spiral arms. In contrast, the late time snapshot in Figure \ref{rvr-large720} has experienced several satellite impacts and the disc contains significant non-axisymmetric structure (even if the disc remains hot). The Milky Way has experienced multiple pericentric passages from the Sagittarius dwarf galaxy, and contains a central bar and spiral arms. Thus, we would expect the response in the \rvr\ plane in the data to closer resemble the late time snapshot, which it does.

\subsubsection{Selection function and biases}\label{bias}
\begin{figure}
\centering
\includegraphics[width=\hsize]{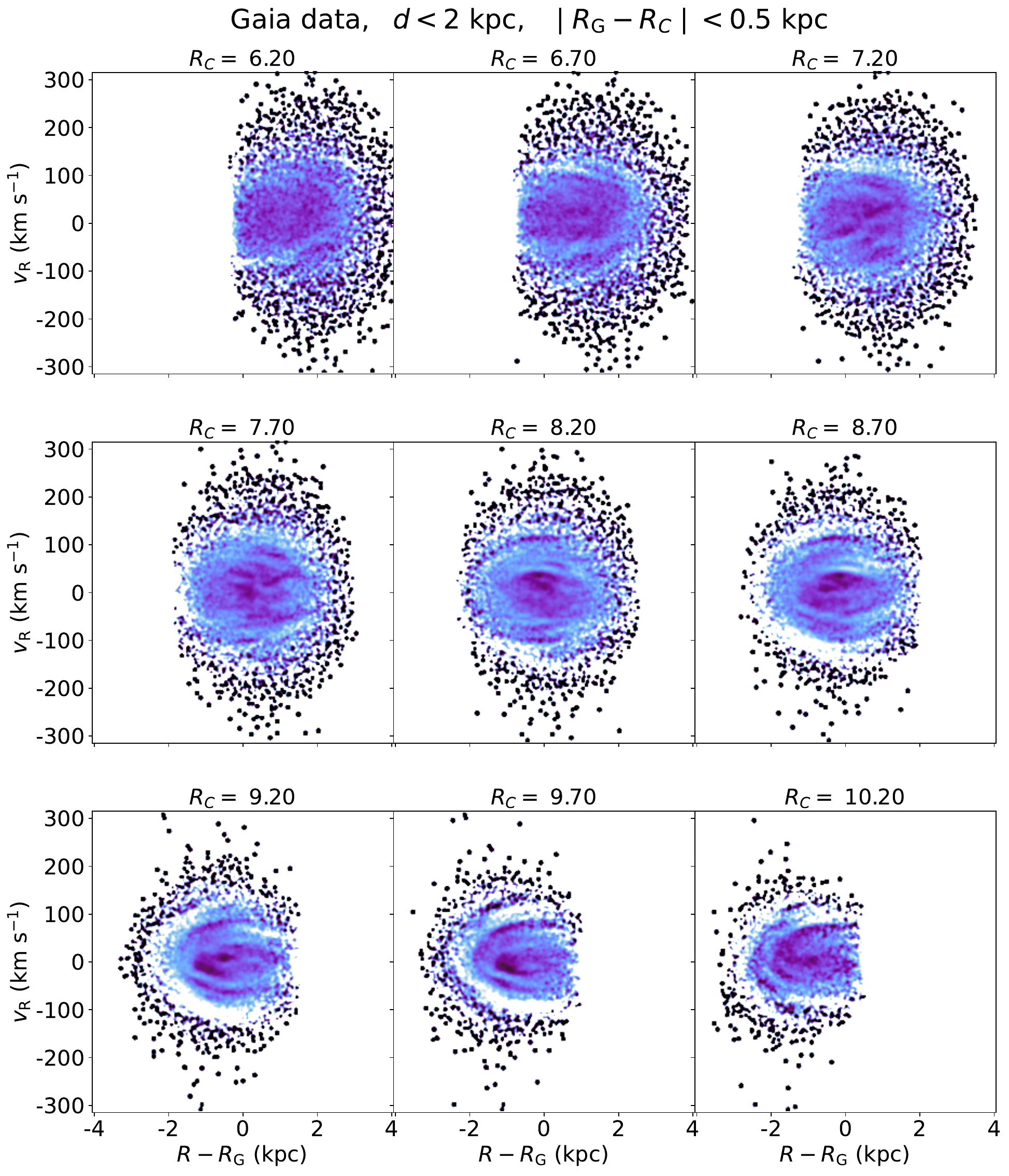}
\caption{Substructure in the radial phase space, $\Delta R-v_{\mathrm{R}}$, for $Gaia$ data within $d<2$ kpc of the Sun and in nine bins of Guiding radius. In each bin, we select stars with $\mid R_{\mathrm{G}}-R_{\mathrm{C}}\mid<0.5$ kpc. Note that this figure has an implicit radial action and radial angle dependence, because stars will only appear if they are at the appropriate part of their radial epicycle. The physical selection limits mean that for stars with low $L_z$ we only see them close to apocentre (e.g., the top left panel), and for stars with high $L_z$ we only see them close to their pericentre (e.g., the lower right panel).}
\label{data-split}
\end{figure}

Note that Figure \ref{data-split} also has an implicit bias in radial action, $J_{\mathrm{R}}$, and conjugate radial angle, $\theta_{\mathrm{R}}$, because stars can either be close to their guiding radius by having low radial action (and are `always' close) while stars with high $J_{\mathrm{R}}$ will only appear if they are at the appropriate part of their radial epicycle. The physical selection limits mean that for stars with low $L_z$ we only see them close to apocentre (e.g. the top left panel), and for stars with high $L_z$ we only see them close to their pericentre (e.g. the lower right panel). Thus across the range of guiding radius bins we probe all radial phases, but for different stars with different $J_{\mathrm{R}}$ and $\theta_{\mathrm{R}}$ and different \rvr\ planes. 

To examine the selection effects and $J_{\mathrm{R}}$, $\theta_{\mathrm{R}}$ biases, we first apply the same selection ($d<2$ kpc, and nine bins in Guiding radius, where $\mid R_{\mathrm{G}}-R_{\mathrm{C}}\mid<0.5$ kpc) to the idealised toy model with a 50 km s$^{-1}$ kick from Section \ref{toy-model}, as shown in the middle row of Figure \ref{toy}. We resample $10^7$ particles within a range of $1144<J_{\phi}<2738$ kpc km s$^{-1}$, to cover a larger range of $R_{\mathrm{G}}$ ($4.8\lesssim R_{\mathrm{G}} \lesssim 11.4$ kpc), but we do not include the effect of dust extinction or $Gaia$ observational uncertainties. 

\begin{figure}
\centering
\includegraphics[width=\hsize]{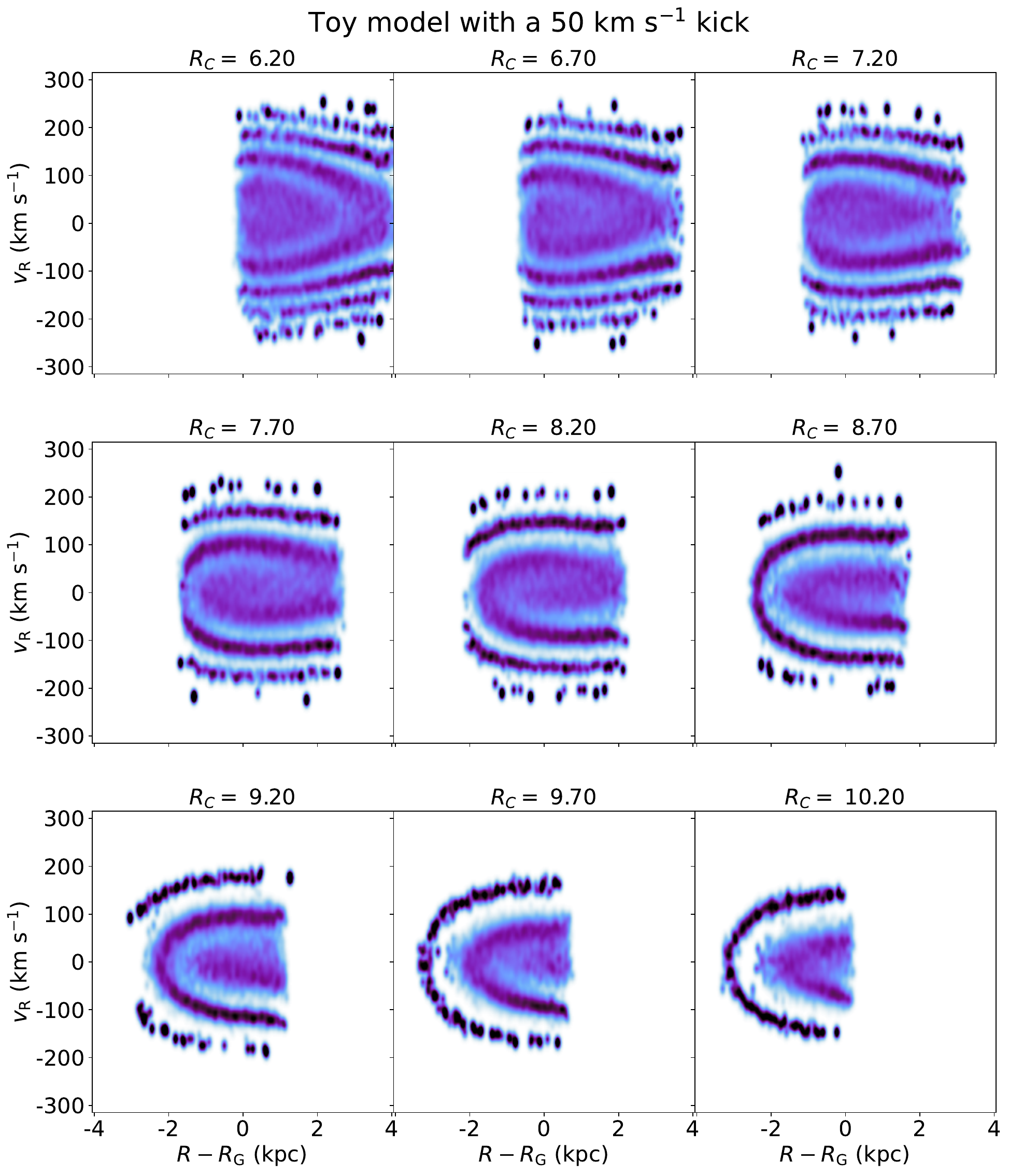}
\caption{Same as Figure \ref{data-split} but for the toy model shown in Figure \ref{toy} with a 50 km s$^{-1}$ kick, after 1 Gyr. Note that the toy does not contain observational uncertainties, selection effects from extinction, or secular effects from a galactic bar or spiral arms.}
\label{model-split}
\end{figure}

Figure \ref{model-split} shows the same angular momentum split as Figure \ref{data-split} but for the toy model. The similarity for higher $J_{\mathrm{R}}$ stars is striking, in that the arcs in the middle and lower rows closely resemble the data, with the middle rows capturing only the arcs at high $v_{\mathrm{R}}$ and low $\mid\Delta R\mid$, for stars close to their Guiding radius, while the lower rows capture the arcs at negative $\Delta R$, from stars at pericentre. The top row's of Figures \ref{model-local} and \ref{data-split} showing the stars at apocentre are also similar, with the flattening of the distributions towards a point at higher $\Delta R$, as expected from Figure \ref{toy}, although the model contains more arcs at high $J_{\mathrm{R}}$ than are recovered in the data.

We then repeat the exercise with the $N$-body model, while also including the effect of dust extinction and $Gaia$ observational uncertainties. We use the stellar population synthesis code \texttt{Snapdragons}\footnote{https://github.com/JASHunt/Snapdragons}\ \citep{Hunt+15,Hunt_snapdragons} to construct mock $Gaia$ data from the $N$-body particles. Because M1 is a pure $N$-body simulation and the particles do not have a self-consistent age or metallicity, we first assign a simple random age and metallicity to each particle, $1<\tau<12$ Gyr and $0.0019<Z<0.019$. This will not result in a realistic spread of stellar populations, nor a realistic age or metallicity gradient across the galaxy. However, it should be sufficient for this simple illustration of the effects of extinction and error.

We then use \texttt{Snapdragons}\ to sample stars from the $N$-body particles using the Padova isochrones \citep{Marigo08_isocrones}, and a Kroupa IMF \citep{Kroupa01_imf}, and convolve them with extinction calculated from a 3D version of the extinction map from \cite{SFD98} and $Gaia$ errors based on 34 months of observations \citep[see][for a full explanation of the mock data synthesis process]{Hunt+15,Grand+Aurigaia}. We then select stars with a fractional parallax error of less than 10 percent, and a line-of-sight velocity error of less than 5 km s$^{-1}$ to match our $Gaia$ sample. This leaves us with a sample of 17,790,885 stars drawn from the simulation, compared to the 16,673,097 which were selected from the $Gaia$ data.

\begin{figure}
\includegraphics[width=\hsize]{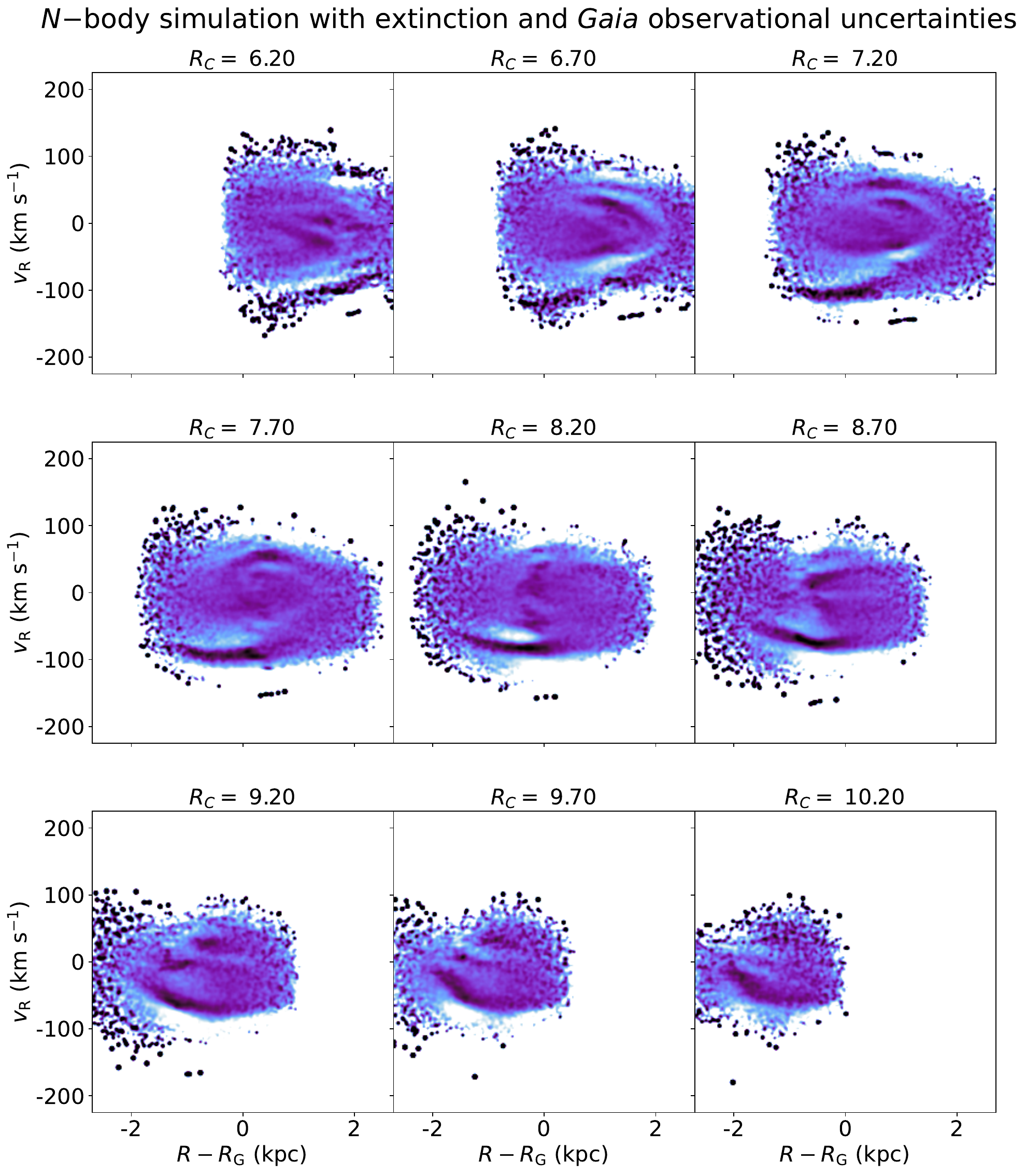}
\caption{Same as Figure \ref{data-split} but for the $N$-body simulation M1, after splitting the $N$-body particles into stars and adding extinction and $Gaia$ observational uncertainties with the population synthesis code \texttt{Snapdragons}.}
\label{M1-split}
\end{figure}

Figure \ref{M1-split} shows the same as Figure \ref{data-split} but for the mock data (selected again with $d<2$ kpc, and nine bins in Guiding radius, where $\mid R_{\mathrm{G}}-R_{\mathrm{C}}\mid<0.5$ kpc) generated from model M1 with \texttt{Snapdragons}. We would not expect it to quantitatively reproduce the $Gaia$ data, as again, this is not a model of the Milky Way, and has not experienced the same perturbation history. However, we recover a clear phase spiral in the upper right and middle left angular momentum bins, and fragments of spirals, in the other bins, qualitatively similar to the $Gaia$ data (although the lower row, with $R_{\mathrm{G}}$ outside the Solar radius only contain arcs at negative $v_{\mathrm{R}}$, unlike the data).

\begin{figure}
\centering
\includegraphics[width=\hsize]{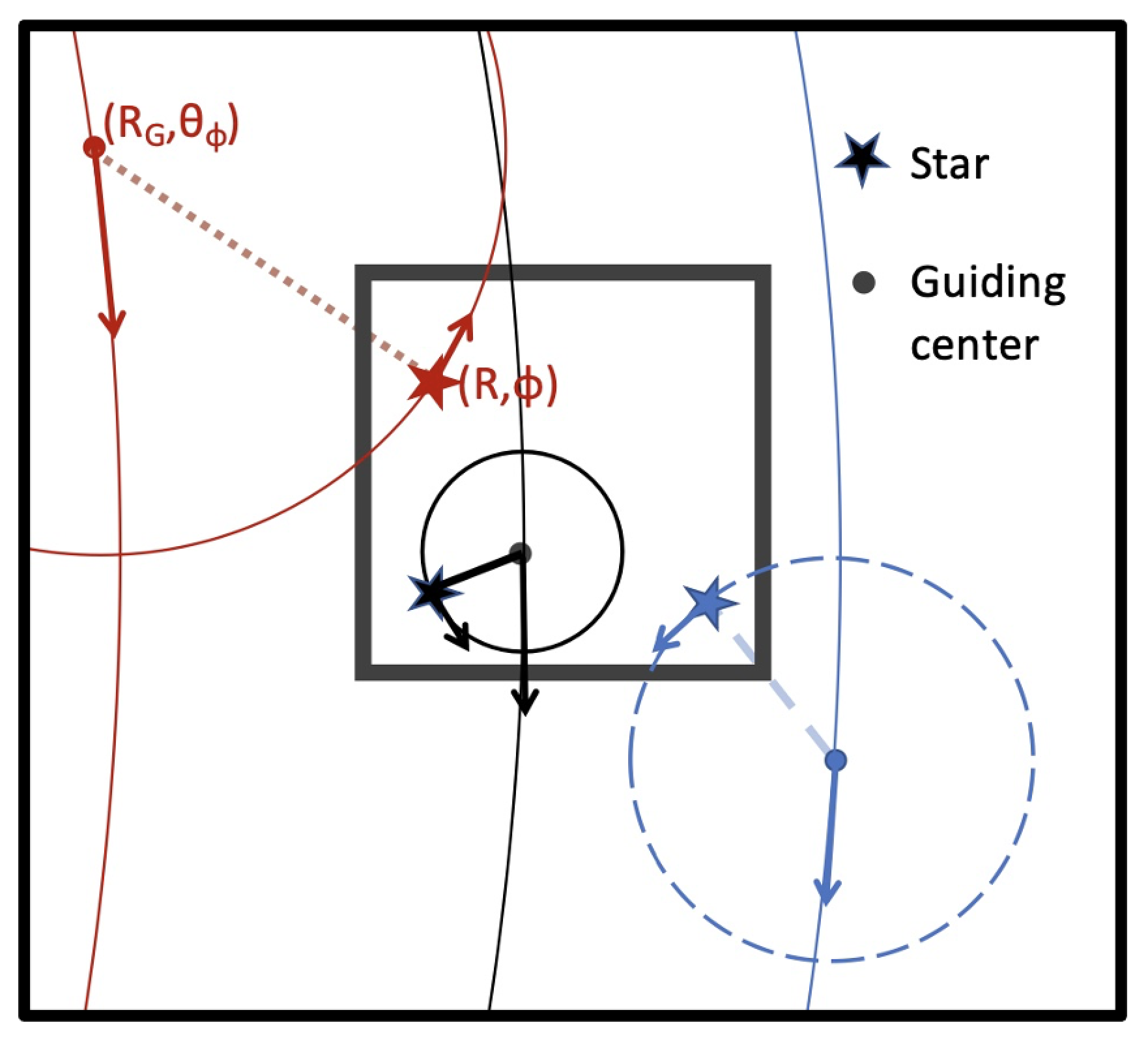}
\caption{Illustration of three stars that are currently in an observed sample (black box) at physical positions $(R, \phi)$ in the disk (star), with their radial epicycles and guiding centres $(R_{\mathrm{G}}, \theta_{\phi})$ shown (dots). The diagram shows that there are biases in radial action, $J_{\mathrm{R}}$, and radial phase, $\theta_{\mathrm{R}}$ as a function of azimuthal action, $J_{\phi}$, and phase $\theta_{\phi}$, when selecting stars in a physically local sample \citep[Reproduced from figure 6 of][]{Darragh-Ford+23}.}
\label{cartoon}
\end{figure}

We further illustrate the phase dependence and bias with the diagram reproduced from Figure 6 of \cite{Darragh-Ford+23}. Figure \ref{cartoon} shows an illustration of three stars that are currently in the observed sample (inner black box) at physical positions $(R, \phi)$ in the disk (star symbols), with their radial epicycles (large circles) and guiding centres $(R_{\mathrm{G}}, \theta_{\phi})$ shown (dots). The `red' star with low angular momentum, $J_{\phi}$, must be on the outer part of its epicycle to appear in the local volume. Conversely, the `blue' star with high $J_{\phi}$ must be on the inner part of its epicycle to appear in the local volume. In addition, stars with their guiding centres further ahead (blue) or behind (red) the Solar neighborhood must have higher radial action, $J_{\mathrm{R}}$, and a specific radial phase angle, $\theta_{\mathrm{R}}$, to appear in the local volume. A star with its guiding centre within the local volume (black) must have sufficiently small $J_{\mathrm{R}}$ to be physically located in the local volume \citep[see Figure 6 and ascociated discussion in][]{Darragh-Ford+23}.

While the selection effects prevent direct observation of a complete \rvr\ phase spiral in the data, the fact that the arc fragments are similar in Figures \ref{data-split},  \ref{model-split} and \ref{M1-split}, and we know that the features in the toy model in Figure \ref{model-split} are parts of coherent \rvr\ phase spirals as shown in Figure \ref{toy}, we consider it likely that the arcs in the data are part of partially observed \rvr\ phase spirals, arising from ongoing radial phase mixing, that will be fully resolvable with future surveys.

However, note that the data shows much finer substructure than either the $N$-body simulation or the toy model, especially at low $J_{\mathrm{R}}$, where Figure \ref{data-split} and Figure \ref{model-split} are significantly different. This is owing to a combination of factors. Firstly, both the simulation and the toy model are limited in resolution which makes it difficult to resolve fine grained phase space structure. Secondly, as mentioned above, the disc in \cite{Hunt+21} is `hot' with a Toomre $\mathcal{Q}$ parameter of 2.2 which suppresses substructure formation while the toy model lacks the self gravity to form self-consistent substructure at all. We would expect bar resonances and spiral arms to produce kinematic substructure at low $J_{\mathrm{R}}$ \citep[such as the classical moving groups, see e.g.][]{Trick+19,STCCR19,Hunt+19} which are not included in Figure \ref{model-split} \citep[note that bar resonances also induce kinematic structure at high $J_{\mathrm{R}}$, e.g.][]{Kawata+21}. Thirdly, the simulation is a pure $N$-body model without gas or star formation, which are also not included in the toy model. The real Milky Way has continued to form stars at low $J_{\mathrm{R}}$, replenishing the inner parts of the \rvr\ plane after earlier perturbations excite older stars to higher $J_{\mathrm{R}}$ \citep[although M1 does contain some low velocity moving groups in Figure \ref{M1-split}, as also shown in $v_{\mathrm{R}}-v_{\phi}$ in][]{Hunt+21}. 

\subsubsection{Utility and context of this space}\label{utility}
As discussed above, the \zvz\ phase spirals are of interest because they allow us to directly see dynamics at work locally in the Galactic disc. Whether phase mixing results from a single perturbation such as Sagittarius, the response to a triaxial dark matter halo, or stochastic scattering, the orbital tracks in \zvz\ allow us to `see' the potential in a similar fashion to how streams from dissolving clusters or dwarf galaxies shows us their orbits in the Milky Way. If we can `see' orbits we can constrain the Galactic potential \citep[e.g. Orbital Torus Imaging;][]{Price-Whelan+21,Horta+23}.

Whether the \rvr\ planes in the data are showing partial phase spirals or simply arcs, they still illustrate the orbital tracks and contain information on the radial gradient of the Galactic potential. For example, the smooth background in \rvr\ can be exploited with a variation of Orbital Torus Imaging \citep{Price-Whelan+21} to measure the radial profile of the Galactic potential, while the pitch angle and amplitude of the \rvr\ phase spiral fragments can be used to `time' the perturbation in the same fashion as being done in studies of the \zvz\ phase spirals \citep[e.g.][]{Antoja23,Frankel23,Darragh-Ford+23}. 

If the \zvz\ phase spirals and the \rvr\ phase spirals originate from the same perturbation, combining the information from both should be more discriminatory than either dimension alone.

\section{Summary}\label{conc}
In this work we explore the radial \rvr\ equivalent of the well known \zvz\ phase spirals. Our conclusions are as follows:
\begin{itemize}
    \item Using a toy model and a simulation of a dwarf galaxy merging into a disc galaxy \citep{Hunt+21}, we show that radial phase mixing of disc stars following a perturbation induces spirals and arc-like substructure in the \rvr\ plane, analogous to the well known \zvz\ phase spirals. The radial frequency of stars, $\Omega_{\mathrm{R}}$, is a much weaker function of radial action, $J_{\mathrm{R}}$, than the vertical frequency $\Omega_z$ is as a function of vertical action, $J_z$. The \rvr\ phase spirals therefore take much longer to phase mix away (see Figure \ref{model-local}).
    \item We present similar substructure in the $Gaia$ DR3 data. In a sample of stars in the very local solar neighbourhood (i.e. within $d < 200~\mathrm{pc}$ of the Sun), the \rvr\ distribution is dominated by the solar neighborhood moving groups, which are local manifestations of extended dynamical structures. However, when exploring a large volume with discrete angular momentum bins we recover numerous arcs which resemble the substructure in the simulation and the toy model. 
    \item Owing to observational selection effects we recover only partial \rvr\ planes for varying angular momentum bins, making it unclear if they are partial recovery of coherent phase spirals or independent arcs. However, applying the same selection to the toy model of \rvr\ phase spirals qualitatively matches the structure observed in the data. 
    \item Regardless, these arcs and ridges follow the approximate orbital paths of stars on their radial epicycles providing us an opportunity to directly observe dynamics at work in the Solar neighborhood, and constrain the Galactic potential in the radial direction \citep[e.g.][]{Price-Whelan+21}.
    \item While we discuss the results in the context of a satellite induced perturbation, any radial perturbation which takes the stellar distribution out of equilibrium will of course result in radial phase mixing, and an \rvr\ phase spiral. As discussed in the introduction, there are many currently proposed theories which can explain the \zvz\ phase spirals; e.g. the Sagittarius dwarf galaxy, buckling of the Milky Way bar, transient spiral structure, a triaxial/lumpy dark matter halo, or stochastic perturbation of the disc by a population of dark matter subhaloes. Such explanations will cause different features in the radial kinematics of stars, and if we can build models of such phase mixing which are coupled in \zvz\ and \rvr, we may be able to break the degeneracy between competing theories, but we defer this to future work. 
\end{itemize}
\vspace{-4pt}

Future data releases from $Gaia$, complemented by spectroscopic surveys such as SDSS-V Milky Way Mapper \citep{Kollmeier+19} will increase the number of stars for which we have 6 dimensional phase space information, and extend the coverage of this data across a larger fraction of the Galactic disc. This may enable us to map full \rvr\ phase spirals, and apply methods to `unwind' them similar to recent work on the \zvz\ phase spirals \citep{Antoja23,Frankel23,Darragh-Ford+23}. For the moment, it may still be possible to fit models to these partial \rvr\ phase spiral fragments \citep[e.g. as done for \zvz\ in][]{WidmarkIII}, or to use the background \rvr\ distribution to constrain the radial component of the Galactic potential \citep[similar to Orbital Torus Imaging as presented in][]{Price-Whelan+21,Horta+23}. 

\section*{Acknowledgements}
We thank the anonymous referee for their constructive comments. We also thank Elise Darragh-Ford \& Kiyan Tavangar for interesting discussion around this work. We make use of data from the European Space Agency (ESA) mission \textit{Gaia} (\url{http://www.cosmos.esa.int/gaia}), processed
by the \textit{Gaia} Data Processing and Analysis Consortium (DPAC;
\url{http://www.cosmos.esa.int/web/gaia/dpac/consortium}). Funding for the DPAC has been provided by national institutions, in particular the institutions participating in the \textit{Gaia} Multilateral Agreement. This research makes use of the SciServer science platform \citep[\url{www.sciserver.org};][]{sciserver}. SciServer is a collaborative research environment for large-scale data-driven science (originally funded by NSF DIBBS). It is being developed at, and administered by, the Institute for Data Intensive Engineering and Science at The Johns Hopkins University. This research made use of the scientific Python package \texttt{SciPy}\ \citep{SciPy}, \texttt{astropy}, a community-developed core Python package for Astronomy \citep{astropy:2013,astropy:2018, astropy:2022}, the $N$-body tree code \texttt{Bonsai}\ \citep{Bonsai,Bonsai-242bil}, the visualisation package \texttt{CMasher}\ \citep{cmasher}, the stellar population synthesis code \texttt{Snapdragons}\ \citep{Hunt+15,Hunt_snapdragons} and the galactic dynamics Python packages \texttt{galpy} \citep{B15}, \texttt{Gala} \citep{gala} and \texttt{Agama} \citep{agama}. KVJ's contributions were supported by NSF grant AST-1715582.
\vspace{-4pt}

\section*{Data availability}
The $Gaia$ data is available at {\url{https://gea.esac.esa.int/archive/}. The full time evolution of model M1 is available on \url{https://sciserver.org/} as part of the SMUDGE data set along with data access tutorials and computing capability.
\vspace{-4pt}
\bibliographystyle{mn2e}
\bibliography{ref2}

\label{lastpage}
\end{document}